# Effect of grain boundary character on segregation-induced structural transitions

Zhiliang Pan, Timothy J. Rupert[*]

Mechanical and Aerospace Engineering, University of California, Irvine, California 92697, USA

*To whom correspondence should be addressed: trupert@uci.edu

## ABSTRACT

Segregation-induced structural transitions in metallic grain boundaries are studied with hybrid atomistic Monte Carlo/molecular dynamics simulations using Cu-Zr as a model system, with a specific emphasis on understanding the effect of grain boundary character. With increasing global composition, the six grain boundary types chosen for this study first form ordered complexions, with the local segregation pattern depending on the grain boundary core structure, then transform into disordered complexions when the grain boundary composition reaches a critical value that is temperature dependent. The tendency for this transition to a disordered interfacial structure consistently depends on the relative solute excess, instead of the grain boundary energy or misorientation angle. Grain boundaries with high relative solute excess go through gradual disordering transitions, whereas those with low relative solute excess remain ordered until high global Zr concentrations but then abruptly transform into thick disordered films. The results presented here provide a clear picture of the effect of interface character on both dopant segregation patterns and disordered intergranular film formation, showing that all grain boundaries are not equal when discussing complexion transitions.

# I. INTRODUCTION

Grain boundary complexions are interfacial structures that are in thermodynamic equilibrium with the abutting crystals [1]. Dillon et al. [2] catalogued six discrete grain boundary complexion types based on their width: (I) a single layer of dopants, (II) clean grain boundaries, (III) bilayers, (IV) multilayers (i.e., more than two layers), (V) intergranular films of nanoscale equilibrium thickness, and (VI) wetting films. Complexions I–V have thicknesses determined by interfacial thermodynamics and can only be stabilized when sandwiched between two crystals, whereas complexion VI is in fact a new bulk phase with arbitrary thickness that is sandwiched between two new interfaces (between the crystals and the new film). The level of structural disorder in general increases from low to high complexion type [2]. Complexions V and VI are often structurally disordered, in which case they can be broadly grouped by the term "disordered intergranular films." These interfacial films are thought to be responsible for activated sintering in ceramics [3-5] and refractory metals [6,7]. Similarly, the change of grain boundary mobility resulting from transitions between different complexion types can be used to understand abnormal grain growth in some materials [2,8-10]. Recent research has also shown that grain boundary complexions can alter the mechanical behavior of materials. For example, bilayer interfacial segregation of Bi can turn normally ductile metals like Cu [11] and Ni [12] into brittle materials. On the other hand, disordered intergranular films may potentially improve the ductility of nanocrystalline materials by facilitating grain boundary sliding [13,14] and by efficiently suppressing crack nucleation and growth induced due to repeated dislocation adsorption at grain boundaries [15,16]. Since complexions offer interesting opportunities for materials design and optimization, the thermodynamics associated with complexion transitions are of great interest.

Complexion transitions occur when there is a discontinuity in the derivative of grain boundary energy with respect to thermodynamic parameters such as temperature and pressure [1]. It therefore follows that the initial grain boundary energy and character prior to doping should be an important factor that affects complexion formation [17]. One kind of important complexion transition is the disordering transition, or a transition from an ordered complexion type to a structurally disordered one. The transition from an ordered interface to complexion type VI would be a traditional melting process, where the bulk liquid phase happens to nucleate at the grain boundary. On the other hand, a transition to complexion V can occur below the bulk melting temperature and is known as "grain boundary premelting" [18]. Experimental work on ceramic systems has indicated that high grain boundary energy tends to promote transitions from ordered to disordered states [2,19-21]. This effect of initial grain boundary energy on complexion transitions has also been predicted from a number of thermodynamic models developed by Tang et al. [22], Luo [23], and Mishin et al. [24].

However, thermodynamic models inherently describe the statistically averaged behavior of structural transitions without considering local details of grain boundary character. In contrast, atomistic simulations can provide detailed information about complexion transitions at an atomic level and can incorporate all of the aspects of grain boundary character. This advantage is especially useful for studying the lower-type, ordered complexions, which are inherently atomistic in nature. For example, recent molecular dynamics simulations uncovered a congruent transition from normal-kite to split-kite shaped grain boundary structural units in face centered cubic metals such as Cu, Ag, Au, and Ni [25]. This subtle structural transition, which would be extremely difficult to observe in experiments or model using thermodynamic theories, successfully explains the unusual, non-Arrhenius behavior of Ag atom diffusion in a grain boundary of Cu [26]. Even for higher-type complexions such as disordered intergranular films, atomistic simulations can provide a more profound understanding. For example, molecular dynamics simulations showed that stable disordered intergranular films in $Si_3N_4$-$SiO_2$ ceramics can decrease the excess energy of the system by reducing the number of dangling bonds at interfaces [27].

A number of atomistic simulations have been performed on complexion transitions in pure materials [18,28-30]. However, studies of segregation-induced structural transitions in alloys are rare. Using an atomistic Monte Carlo method in a semi-grand canonical ensemble [31], Williams and Mishin [32] studied the gradual premelting of a Σ5 grain boundary in Cu-Ag alloy system. However, the effect of grain boundary character on segregation-induced complexion transitions has not been investigated to date, even though these details should greatly impact complexion formation. In this work, hybrid atomistic Monte Carlo/molecular dynamics simulations are used to study the effect of grain boundary character on segregation-induced complexion transitions in Cu-Zr binary alloys, with the equilibrium chemical distribution of Zr atoms in Cu matrix modeled with Monte Carlo methods and structural relaxations captured by molecular dynamics. The six grain boundaries studied here, which cover various aspects of boundary character, go through transitions from ordered grain boundaries to disordered intergranular films either gradually or abruptly when the grain boundary composition resulting from the segregation process reaches a critical value. Rather than grain boundary energy, a more consistent indicator for the ease of the disordering transition is the relative solute excess [33] that represents the grain boundary's ability to adsorb Zr atoms. This work directly demonstrates that grain boundary segregation is crucial for inducing structural transitions, with dopant adsorption ability playing a decisive role.

## II. COMPUTATIONAL METHODS

Hybrid Monte Carlo/molecular dynamics simulations were performed using the Large-scale Atomic/Molecular Massively Parallel Simulator (LAMMPS) code [34], with all molecular dynamics simulations using a 1 fs integration time step. Cu-Zr was chosen as a model alloy system here to study segregation-induced complexion transitions because both Cu and Zr are transition metals and Zr should segregate to the grain boundaries due to its low solid solubility in Cu [35]. In addition, the good glass-forming ability of Cu-Zr alloys [36] makes it highly likely that disordered intergranular films can be accessed in this alloy system. Embedded-atom method potentials were used to describe the Cu-Cu and Zr-Zr interactions, while a Finnis-Sinclair potential was used to describe the interactions between Cu and Zr atoms [37]. All structural analysis and visualization of atomic configurations was performed using the open-source visualization tool OVITO [38]. The local crystal structure of each atom was identified based on common neighbor analysis (CNA) [39], with face centered cubic atoms colored green, hexagonal close packed atoms red, body centered cubic atoms purple, icosahedral atoms yellow, and other atoms white. Zr atoms are colored blue to highlight the dopants.

Six bicrystal samples with different grain boundary character were used as starting configurations. As shown in Fig. 1, three symmetric tilt grain boundaries and three twist boundaries were created with different misorientation angles and grain boundary plane normals, and thus different grain boundary energies. Of the six grain boundaries, the Σ5 (013) interface shown in Fig. 1(a) has the highest grain boundary energy of 0.062 eV/Å$^2$ at 0 K, with a kite-shaped core containing E structural unit [40]. On the other hand, the Σ3 (111) twin boundary shown in Fig. 1(b) has the lowest energy of 0.001 eV/Å$^2$. The Σ11 (113) boundary shown in Fig. 1(c) has an intermediate grain boundary energy of 0.022 eV/Å$^2$, with C structural units [41,42]. The twist

(100) 36.86º boundary (0.056 eV/Å$^2$) shown in Fig. 1(d) and the twist (100) 10.39º interface (0.034 eV/Å$^2$) shown in Fig. 1(e) share the same (100) grain boundary plane but have different twist angles, to study the effect of misorientation angle. Finally, the twist (111) 30º boundary (0.027 eV/Å$^2$) shown in Fig. 1(f) has the same (111) grain boundary plane orientation as the Σ3 (111) boundary, but has a misorientation angle of 30º that is only half of the value associated with the twin boundary. When building the twist boundaries, we did not explore all possible translations or insert and remove additional atoms to find the absolute minimum energy state. Recent work from Han et al. [43] has demonstrated that twist grain boundaries experience only very small changes to their energy with such variations.

The six simulation cells are approximately 92 nm long (X-direction), 5–6 nm tall (Y-direction), and 4–6 nm thick (Z-direction) with each containing ~200,000–300,000 atoms. Periodic boundary conditions were applied in all directions. The samples were first equilibrated with a conjugate gradient minimization technique so that the grain boundary structure reaches a stable state with minimum potential energy, from which the grain boundary energies at 0 K were calculated. A Nose-Hoover thermo/barostat was then used to further relax the sample for 200 ps under zero pressure at different temperatures (600–1200 K in increments of 50 K). Thereafter, doping with Zr solutes was simulated using a Monte Carlo method in a variance-constrained semi-grand canonical ensemble [44] after every 100 molecular dynamics steps, with the target global composition of Zr fixed to different values that increase in small increments. It should be noted that in classical semi-grand canonical ensemble Monte Carlo simulations, the global composition is adjusted by changing the chemical potential difference [32]. In the variance-constrained version used here, the global composition is controlled mainly by setting a predetermined target value.

The chemical potential difference is given an initial guess and then adjusted during the simulation to achieve the required global composition [44].

Fig. 2 shows how the potential energy of a bicrystal sample containing Σ5 boundaries changes with the number of Monte Carlo steps. Fig. 2(a) shows that the potential energy of the system first decreases quickly with Monte Carlo iteration as Cu atoms are replaced with Zr atoms, then appears to saturate after a few Monte Carlo steps. However, reducing the limits of the Y-axis, as presented in Fig. 2(b), shows that the potential energy continues to decrease with additional Monte Carlo steps. Finally, the system reaches thermodynamic equilibrium, where only small fluctuations around a constant energy are observed (Fig. 2(c)). Here, the system is considered to be equilibrated when the absolute value of the fitted slope of the potential energy over the last 4000 Monte Carlo steps is less than 0.001 eV/step, since additional Monte Carlo steps lead to no considerable structural changes. A conjugate gradient energy minimization was then used to remove thermal noise, so that the interfacial structure obtained during the doping process can be preserved and only the interfacial regions are identified as defects according to CNA. At this point, the volume of each atom is also calculated based on the Voronoi tesselation. Finally, the average grain boundary thickness is measured by taking the total volume of the defect atoms in the interfacial region and dividing by the cross-sectional area.

## III. RESULTS AND DISCUSSION

Fig. 3 shows the equilibrium structures of the Σ5 (013) grain boundary during the doping process at three temperatures. When the global composition of Zr is 0.1 at. %, dopant atoms occupy the tip of the kite structure at 600 K and 900 K to form a single layer complexion. At 1200

K, the boundary starts to become partially disordered, but the zoomed image shows that some crystalline order still exists and the kite structure can still be observed, albeit with a distorted shape. When the global composition is increased to 0.3 at. % Zr, the grain boundary remains ordered at 900 K while becoming partially disordered at 600 K, with a small section becoming disordered but the majority of the boundary still doped in the single layer pattern. The grain boundary becomes completely disordered and an intergranular film forms at 1200 K. When the global composition is increased to 1.1 at. % Zr, the grain boundary is fully disordered at all three temperatures and the disordered film thickness increases as temperature increases. The radial distribution function of a transformed grain boundary at 1200 K and a global composition of 2.1 at. % Zr is shown in Fig. 4(a). The peaks of this function vanish rapidly with increasing pair separation distance and no clear peaks can be identified after 1 nm, indicating the absence of long-range, crystalline order in this region. The splitting of the second peak signifies that there is some short-range order in this region. The structures that contribute to the short-range order are mostly icosahedral type, as shown in Fig. 4(b). The packing of these icosahedral clusters also indicates very limited medium-range order. Such limited short to medium-range order is typical of the atomic structures of metallic glasses [45]. Our analysis of other doping conditions and temperatures found this same lack of crystalline order and presence of short to medium-range order in all disordered grain boundaries. The disordered intergranular films become thicker when the global composition is further increased to 2.1 at. % Zr.

To provide a complete view of the complexion transition process, grain boundary thickness, grain boundary composition, and grain interior composition are plotted as a function of global composition in Fig. 5. Four different complexion transition stages can be identified and are marked in this figure. The first complexion is an ordered grain boundary, identified as a single layer of dopants for this boundary type. This complexion structure is found only at low global compositions and relatively low temperatures. As shown in the inset to Fig. 5(a), the grain boundary thickness in this stage does not change with global composition and is about 0.5 nm, or roughly the thickness of the initial grain boundary. During this stage, the grain boundary and grain interior compositions increase quickly with global composition, as shown in Fig. 5(b) and (c). The second region is a gradual transition stage where the grain boundary is only partially disordered, located between the dotted and dashed lines in Fig. 5(a) and (b). During this transition stage, more atoms near the interface are identified as defect atoms and the grain boundary thickness increases with the global composition. The increase of the volume fraction of grain boundary atoms outpaces the increase of the number of Zr atoms at a global composition of ~0.5 at. % Zr, leading to a slight decrease in the grain boundary composition at some temperatures in Fig. 5(b).

The disordering process continues with increasing global composition until the grain boundary becomes completely disordered and nanoscale intergranular films have formed. The grain boundary thickness is approximately 1 nm when this complexion structure arises. The grain boundary composition only increases slowly with global composition in this stage, finally saturating to constant compositions as denoted by the horizontal dotted lines in Fig. 5(b), where wetting films, or complexions of Type VI, finally form at the grain boundary. Further increasing the global composition to the maximum explored value of 4 at. % Zr for this interface does not change the saturated trend. The thick solid lines in Fig. 5(a) and (b) mark where grain boundary composition saturates and thus the formation of a wetting film when the boundary reaches a thickness of about 4 nm. Nanoscale intergranular films form due to the interaction between two disordered-crystalline interfaces that sandwich the disordered film, which can be characterized by a disjoining potential [46]. Past work [18,24,46,47] which calculated this potential for different metals suggests that a spacing on the order of a few nanometers can eliminate the interfacial interaction, with ~4 nm apparently being large enough to do so for Cu-Zr. Because the wetting film has a set Zr concentration for a given temperature, the volume fraction and therefore the thickness of the wetting film increases linearly with increasing global composition.

To observe how boundary character can affect complexion transitions, we next investigated the doping behavior of the Σ11 (113) grain boundary. Fig. 6 shows the equilibrium structures of this interface during the doping process. When the global composition is 0.3 at. % Zr, the grain boundaries are decorated with Zr atoms in an ordered manner at all three temperatures. The ordered segregation pattern is different as well, with all corners of the C structural unit being possible adsorption sites, although the three possible sites are not fully occupied by Zr atoms. When the global composition is increased to 0.9 at. % Zr, the grain boundary at 600 K becomes fully disordered, but still remains ordered at 900 K and 1200 K. Increasing the global composition to 2.1 at. % Zr leads to the formation of disordered intergranular films at all three temperatures. This transition behavior differs from the Σ5 (013) grain boundary, where the disordering transition occurred at a much lower global composition and at the highest temperature of 1200 K first.

Plots of the grain boundary thickness, grain boundary composition, and grain interior composition as a function of the global composition are shown in Fig. 7 for the Σ11 (113) boundary, where three transition stages can be identified for the 600 K simulation set. The grain boundary has an ordered structure in a multilayer (trilayer) segregation mode when the global composition is less than 0.5 at. % Zr, with a constant grain boundary thickness less than 0.5 nm. The premelting transition occurs when the grain boundary becomes disordered and nanoscale intergranular films form at global compositions greater than ~0.5 at. % Zr. The grain boundary composition increases slowly as the global composition increases in this stage and then finally saturates to an equilibrium concentration level where the wetting film finally forms. The grain interior composition also saturates to an equilibrium concentration level. At 900 K and 1200 K, only two stages can be identified. As shown in Fig. 7(a), the grain boundary thickness first does not change as the global composition increases, then jumps to a fairly large value above ~1.3 at. % Zr. The jump of the specific volume (volume per atomic mass unit) of the grain boundary around this global composition, as shown in the inset to Fig. 7(a), indicates an abrupt or first-order transition without any intermediate states. After increasing almost linearly with the global composition before the abrupt transition, the grain boundary and grain interior compositions shown in Fig. 7(b) and (c), respectively, also abruptly shift to a saturated value, confirming the first-order nature of the transition. Since the grain boundary composition jumps to the equilibrium concentration of a disordered structure, this process indicates the formation of wetting films. The observation of abrupt disordering transitions is dramatically different from the gradual complexion transition behavior that was observed for the Σ5 (013) grain boundary at all temperatures.

To isolate the abrupt disordering transition at 900 K and 1200 K, nine additional simulations were performed with target global compositions between 1.22 and 1.38 at. % Zr, separated by very small increments. The results in Fig. 7 show that the grain boundary thickness and composition either stay as ordered grain boundaries or transform into wetting films, with no intermediate states between them. Note that at 900 K the abrupt transition occurs at a global composition of 1.34 at. % Zr, but does not at a slightly higher value of 1.35 at. % Zr. This is because a first order disordering transition may go through a myriad of possible reaction paths that are thermodynamically identical but have slightly different energy barrier. Due to these small path-to-path variances, it is possible that one reaction path has higher energy barrier than another even if the global composition of the former is slightly higher than that of the latter. Nevertheless, the difference between the energy barriers of these reaction paths is minute, explaining why this abnormal behavior only happens in a small window of global composition.

Fig. 8 shows the equilibrium structures of the twist (111) 30º grain boundary during the doping process. The grain boundary remains ordered at the three temperatures to a global composition as high as 1.1 at. % Zr, far later than the composition where both the Σ5 (013) and the Σ11 (113) grain boundaries had already become disordered, and Zr atoms are segregated into a bilayer pattern. With increasing global composition, the grain boundary transforms into a disordered intergranular film first at 600 K and then at 1200 K, but remains ordered at 900 K even at a global composition of 3.1 at. % Zr. Plots of grain boundary thickness, grain boundary composition, and grain interior composition are shown in Fig. 9, where three doping stages can be identified at 600 K. The grain boundary remains ordered up to a global composition of ~1.44 at. % Zr, then transforms into a nanoscale intergranular film. Finally, a wetting film forms at a global composition of ~2.24 at. % Zr, as indicated by the saturated grain boundary composition. Only two doping stages can be identified at 1200 K. The grain boundary first remains ordered up to a global composition above ~2 at. % Zr, then transforms into a disordered wetting film abruptly. Again, additional simulations around this critical composition showed no intermediate states. At 900 K, only one complexion type was identified, with the grain boundary remaining ordered in a bilayer segregation mode throughout the entire doping process up to the maximum explored composition of 4 at. % Zr. The disordering transition is thus expected to occur at higher global compositions. The grain boundary thickness remains constant, and both the grain boundary and grain interior compositions increase gradually with increasing global composition.

Fig. 10 presents the equilibrium structures of the remaining three grain boundary types at 1200 K and a global composition of 0.27 at. % Zr. Two grain boundaries remain ordered, but with different segregation patterns. Zr atoms are segregated into the $\Sigma 3$ (111) twin boundary in a single layer mode, whereas the twist (100) 10.39º grain boundary shows a multilayer (four layers) segregation state. The twist (100) 36.86º grain boundary is partially disordered, but crystalline order still exists, as indicated by the legible boundary plane in the upper part of the image. Plots of grain boundary thickness, grain boundary composition, and grain interior composition at 1200 K are shown in Fig. 11, with all six grain boundaries plotted together in this figure. The twist (100) 36.86º grain boundary first gradually transforms into a nanoscale intergranular film and then a wetting film, qualitatively mimicking the behavior of the $\Sigma 5$ (013) grain boundary. Alternatively, the rest of the grain boundaries show abrupt transitions from ordered complexions to disordered wetting films, with the transition of the $\Sigma 3$ (111) twin boundary occurring at the highest global composition. The wetting films (bulk disordered phase) together with the grains (crystalline phase) form a two-phase region, where both phases have fixed compositions at a given temperature and the fraction of each phase is determined according to the lever rule at a given global composition, independent of the grain boundary type. This can be confirmed by the fact that the grain boundary and grain interior compositions of all the samples saturate to the same level. The saturated grain boundary and grain interior compositions are in fact the liquidus and solidus lines, respectively, of an equilibrium phase diagram. The grain boundary and grain interior compositions described throughout this paper are more fundamentally important quantities than global composition values, which will depend on the sample dimensions or grain boundary area per unit volume in the simulation cell. Global compositions are still useful for making qualitative comparisons between samples within this study though.

To clarify the effect of temperature, Fig. 12 presents grain boundary thickness, grain boundary composition, and grain interior composition as a function of temperature while fixing the global composition to 4 at. % Zr. From this figure, three different complexion transition behaviors can be observed. Four grain boundaries form disordered wetting films while the Σ3 (111) twin boundary is an ordered complexion for the entire temperature range. Alternatively, the twist (111) 30° grain boundary forms a disordered wetting film at temperatures below 800 K and above 950 K. At intermediate temperatures, however, it remains ordered. The thicknesses of the disordered interfaces increase with increasing temperature, independent of the grain boundary type. At 600 K, a second crystalline phase is formed during the doping process in the twist (111) 30°, twist (100) 10.39°, and Σ5 (013) bicrystal samples, as shown in the inset to Fig. 12(a). This crystalline phase had a Zr composition of ~18.6 at. %, close to that of the Cu-rich compound $Cu_9Zr_2$ [48]. The formation of a second phase draws a large amount of Zr atoms from the disordered intergranular films, leading to a decrease of the film thickness and increase of the average grain interior composition. These effects are most easily seen for the twist (111) 30° grain boundary, where the largest amount of second phase formed. When the temperature is between 650 and 750 K, there is still a noticeable difference between the twist (111) 30° grain boundary and other boundaries that have gone through disordering transitions, although no second crystalline phase is identified in this temperature range. Prior molecular dynamics simulations of melting of Cu clusters showed that the (111) surface is difficult to melt [49]. This might indicate that the (111) grain boundary plane is also difficult to melt and requires high Zr concentration at boundaries to drive the melting process, leading to slightly higher boundary composition and reduced thickness for the twist (111) 30° grain boundary at temperatures between 650 and 750 K.

Figs. 3–11 show that grain boundary character has a prominent effect on segregation-induced structural transitions. In the first doping stage, boundaries remain ordered and are doped with Zr atoms to form different lower-type complexions. We find that the details of the segregation pattern

are determined by the initial grain boundary structural units. The Σ5 (013) grain boundary with kite-shaped structural units has preferential adsorption sites distributed at the tip of the kite, favoring a single layer segregation mode. The coherent twin boundary has adsorption sites distributed on the single layer twin plane, also giving single layer mode. The twist (111) 30° grain boundary usually has two identical atomic planes inside the core, giving a bilayer segregation mode. The adsorption sites in the Σ11 (113) grain boundary can be any corner of the C structural unit, resulting in a trilayer complexion. The core structure of the twist (100) 10.39° grain boundary is in fact a network of screw dislocations which span four atomic layers, as shown in Fig. 1(e), resulting in a multilayer segregation mode.

Our observations help explain the variety of segregation patterns that have been reported from experiments in the literature, including the single dopant layer observed for a Cu Σ5 (013) grain boundary when either Bi [50] or Ag [26,51] are added, the observed single layer of Zn or Gd for coherent twin boundaries in Mg [52], the bilayer segregation pattern in a Au-doped Si Σ43 twist (111) 15° grain boundary [53], and the trilayer in a Ga-doped Al Σ11 (113) grain boundary [54]. The importance of initial grain boundary structure on segregation patterns was also implied by Frolov and coworkers, who showed that kite shaped Σ5 (013) and filled-kite shaped Σ5 (012) grain boundaries of Cu only have a single layer segregation pattern when doped with Ag, while a split-kite shaped Σ5 (013) boundary demonstrates a bilayer segregation pattern [26,51]. Our work and the existing literature show that grain boundaries of the same character should have similar segregation patterns, regardless of specific material system, suggesting that the segregation mode is determined by atomic boundary structure and the distribution of the potential adsorption sites. Because of the effect of grain boundary structure, complexion transitions at a specific interface may not move from single layer to bilayer and then to trilayer, as predicted from a discrete diffuse-interface model [23]. Transitions between monolayer, bilayer, and trilayer may not be easily observed for a single interface, as we observe that only one type of segregation pattern is stable for each grain boundary type. To make this transition occur, a change of core structure and a redistribution of the potential adsorption sites would be required.

Grain boundary character also affects the critical temperature and Zr concentration for the breakdown of structural order at doped interfaces. Two grain boundaries go through gradual transitions and four others go through abrupt transitions in Fig. 11 as Zr concentration is increased. It has been previously hypothesized that a high starting grain boundary energy promotes continuous or gradual disordering transitions, whereas low grain boundary energy delays these changes and encourages abrupt transitions [23]. In this work, the two grain boundaries that go through the gradual premelting transition do in fact possess the highest energy of the six chosen grain boundaries. To quantify this effect more completely, the global disordering composition, when each grain boundary last has an ordered structure at 1200 K (i.e., the maximum amount of dopant that can be added before the ordered interfacial structure breaks down and either a nanoscale intergranular film or wetting film forms), is plotted as a function of grain boundary energy in Fig. 13(a). In general, the interfaces with higher grain boundary energies do go through disordering transitions at lower global compositions, but an outlier exists. The twist (111) 30° grain boundary has a higher energy than the $\Sigma 11$ (113) interface, but the disordering transition is more difficult to access and requires a higher global Zr composition. The misorientation angle between neighboring grains is not a consistent indicator for the propensity of disordering either. Comparison of the twist (100) 36.86° and twist (100) 10.39° shows that a large misorientation angle correlates with earlier disordering transitions for a (100) boundary plane. Alternatively, comparing the $\Sigma 3$ (111), where the misorientation angle is 60°, and the twist (111) 30° boundaries shows that a large misorientation angle correlates with delayed disordering transitions for a (111) boundary plane, indicating the opposite trend.

Fig. 11(b) shows that, prior to the formation of disordered intergranular films, grain boundary composition increases faster at interfaces with easier disordering transitions. For example, the Zr concentration in the Σ5 (013) grain boundary increases at the highest rate with respect to the global composition, while the Σ3 twin boundary is slowly doped with Zr and transforms into a disordered intergranular film at the highest global composition. This suggests that the rate of Zr incorporation into the grain boundary is a useful indicator of how easily a disordering transition occurs. To test this hypothesis, the global composition at which each interface last has an ordered structure was then plotted as a function of the relative solute excess, defined as the ratio of grain boundary composition divided by grain interior composition [33] in Fig. 13(b). The curve shows that relative solute excess gives a more consistent indicator of how easily a disordered complexion can form, suggesting that the high dopant adsorption ability, rather than the grain boundary energy, leads to the preferential disordering transitions in random high-angle grain boundaries in the polycrystalline samples used in experimental studies in the literature [2,19-21].

Fig. 11(b) also shows that the disordering transition for all of the boundaries studied here occurs at approximately the same grain boundary concentration level (~8 at. % Zr at 1200 K) and does not strongly depend on the boundary type. This is more clearly seen in Fig. 14, which gives the boundary disordering composition (the critical grain boundary composition above which ordered interfacial structure breaks down) as a function of temperature. This value is the maximum amount of Zr a grain boundary can accept through a segregation process while remaining ordered, marking the transition point where more Zr causes disordering. The equilibrium composition of wetting films at different temperature is also shown in Fig. 14 as a solid line, which gives the upper bound on grain boundary composition for a disordered intergranular film. The space between the boundary disordering composition and wetting film composition marks the possible composition range where nanoscale intergranular films can be formed. At 600 K, 900 K, and 1200 K, the $\Sigma 5$ (013) and $\Sigma 11$ (113) grain boundaries go through this transition at almost identical boundary compositions. The data for all six boundary types is also tightly clustered at 1200 K. Our findings here suggest that a variety of complexions may be present in a polycrystalline material with a given global composition and a random collection of grain boundaries. Random high-angle grain boundaries have high interfacial energy and will be rapidly covered with dopant atoms, like the $\Sigma 5$ (013) boundary shown in Figs. 3 and 5, meaning they can easily transform into disordered intergranular films. In contrast, boundaries such as a $\Sigma 3$ twin boundary have very small relative solute excess and would require much higher global compositions to reach the level of boundary doping necessary to induce a disordering transition. Therefore, a random polycrystalline material would likely have multiple complexion types for a given global composition.

A more complete picture of the interplay between dopant segregation and the loss of ordered boundary structure is shown in Fig. 15, which can be considered a "disordering transition diagram." The segregation behavior of an ordered interface can be roughly represented by the black curves corresponding to a Langmuir-McLean isotherm [1,55], with each line having a fixed global composition. The Langmuir-McLean isotherm is a classical grain boundary segregation model that assumes there are a fixed number of identical segregation sites per unit area at a grain boundary and that the material acts as an ideal solution, which requires negligible enthalpy of mixing and thus a dilute grain interior composition [56]. The first assumption is satisfied here since each grain boundary gives a specific distribution of preferential adsorption sites. The second assumption is also approximately satisfied since grain interior composition is relatively low when compared to the grain boundary composition for most of the interfaces studied here, with the notable exception of the $\Sigma 3$ (111) boundary. The only adjustable parameter in the Langmuir-McLean isotherm is the grain boundary segregation energy which can give an approximate linear relationship between grain boundary and grain interior compositions, from which a relative solute excess can be calculated. In this work, the segregation energy was modulated so that the resulted relative solute excess at grain boundaries is close to that of the grain boundaries shown in Fig. 13(b). The boundary disordering composition, which was shown in Fig. 14, is replotted here as a red line. Black curves to the left of the red line correspond to ordered grain boundaries, whereas black curves to the right of the red line correspond to grain boundaries that will go through disordering transitions.

Our results show that the effect of temperature on disordering transitions is two-fold. On one hand, the red curve shows that increasing temperature can promote a disordering transition by reducing the critical boundary disordering composition required for the formation of disordered intergranular films. On the other hand, the segregation curves show that increasing temperature can also suppress such a transition by reducing the grain boundary excess for a given global Zr composition. The competition between the two temperature effects determines the transition to a disordered complexion. It is important to note that the curvature of the disordering curve is convex, while the curvature of each segregation curve is concave. With increasing temperature, the grain boundary composition at a given global composition decreases first faster in a relatively low temperature range, then slower in a relatively high temperature range than the boundary disordering composition. This difference can roughly divide the transition diagram into three different regions. At high temperature, the disordering transition is temperature-dominated and can occur despite weak levels of grain boundary segregation. One extreme case is that grain boundary premelting can occur in pure materials when the temperature is close to the melting point [32,57,58]. At low temperature, the transition can still occur due to the strong dopant segregation and is chemistry-dominated. However, at intermediate temperatures, neither temperature nor chemical concentration dominates and the disordering transition is most difficult. A higher global composition would be required to induce disordering at this temperatures. This behavior is exactly what was observed in Fig. 12 for the twist (111) 30º grain boundary that goes through disordering transition at temperatures below 800 K and above 950 K, but remains ordered between the two temperatures for a global composition of 4 at. % Zr.

Fig. 15 demonstrates the importance of grain boundary segregation for inducing structural transitions, with the simple Langmuir-McLean model giving an approximate description of the segregation behavior at an ordered interface. This viewpoint incorporates the role of dopants into the effect of grain boundary character on the segregation-induced structural transition behaviors, and captures subtleties missed by simply referring to grain boundary energy. Even considering that different dopants may have different segregation behavior and disordering composition due to the different bonding character, it is expected that a general feature shared between them will be that the boundary adsorption ability plays a decisive role for inducing complexion transitions.

## IV. CONCLUSIONS

In this work, hybrid atomistic Monte Carlo/molecular dynamics simulations were performed to study segregation-induced complexion transitions, using Cu-Zr as a model system. Different type of grain boundaries were used to reveal the effect of boundary character. Based on the results shown here, the following conclusions can be drawn:

- Ordered complexions first form at grain boundaries during the doping process, with segregation pattern and complexion type depending on the starting grain boundary structure. This structure determines the distribution of preferential adsorption sites and acts as a template for dopant segregation.

- The transition from ordered complexions to disordered intergranular films can occur either gradually or abruptly when the grain boundary composition reaches a temperature-dependent critical value that is not a strong function of grain boundary character.

- The ability to absorb dopant atoms, indicated by the relative solute excess as opposed to the grain boundary energy, directly determines the propensity for boundary to undergo a disordering transition. Grain boundaries with high adsorption ability tend to go through gradual transitions at low global Zr compositions, whereas those with low adsorption ability tend to go through abrupt or first order transitions at higher global Zr compositions.

- Nanoscale intergranular films tend to form at grain boundaries with high relative solute excess, with compositions between the initial boundary disordering composition and wetting film composition. This nanoscale disordered film will transform into a disordered wetting film eventually as global composition continues to increase. Grain boundaries with low dopant adsorption ability do not favor the formation of nanoscale intergranular films with equilibrium thickness and transition directly to wetting films.

- The interplay between temperature and chemistry effects can divide the disordering transition into three different behaviors: temperature-dominated behavior at high temperatures, chemistry-dominated behavior at low temperatures, and difficult transition behavior at intermediate temperatures.

The results presented here directly show that grain boundary segregation is essential for inducing transitions between complexion types, especially at relatively low temperatures where ordered segregation patterns develop. This work explains the scattered observations of ordered complexions found in the literature and gives a better indicator for the tendency for disordering. Finally, disordering transition diagram explain the interplay between temperature and chemistry and should be useful for predicting complexion transitions.

## ACKNOWLEDGEMENTS

This research was supported by U.S. Department of Energy, Office of Basic Energy Sciences, Materials Science and Engineering Division under Award DE-SC0014232. T.J.R. acknowledges support from a Hellman Fellowship Award.

**FIGURES AND CAPTIONS**

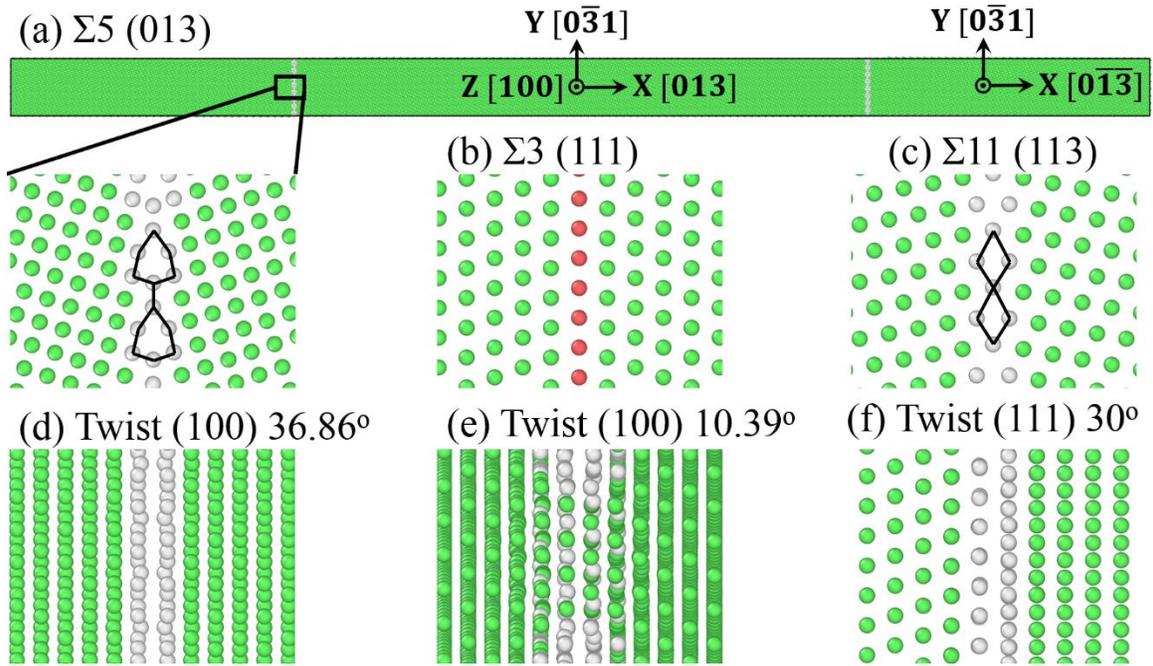

FIG. 1. (a) A starting bicrystal sample containing two Σ5 (013) grain boundaries with kite shaped structural units, shown as a close-up below the sample. The grain boundary structures of the other five bicrystal samples containing (b) Σ3 (111), (c) Σ11 (113) with C structural unit, (d) twist (100) 36.86º, (e) twist (100) 10.39º, and (f) twist (111) 30º grain boundaries are also shown here.

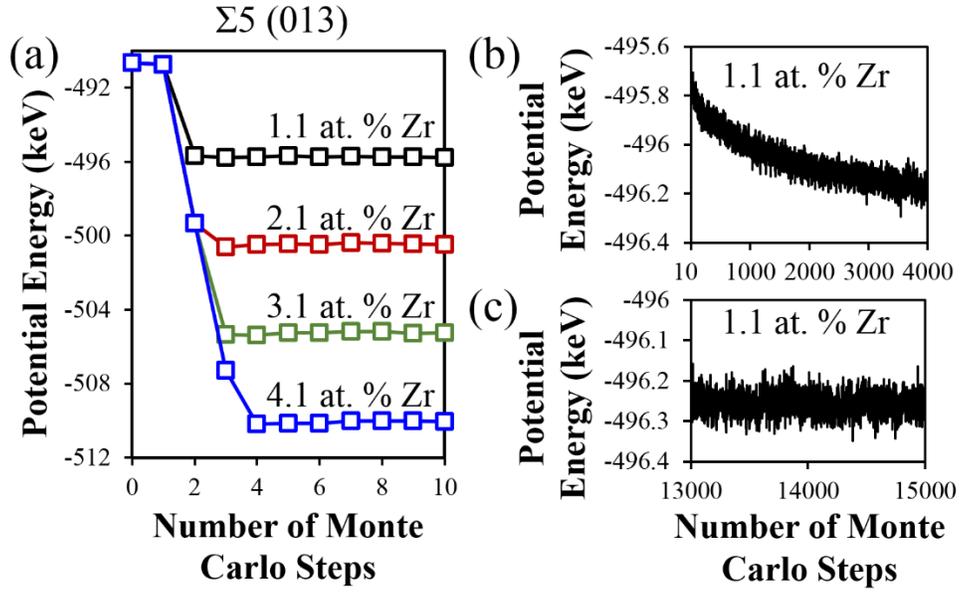

FIG. 2. Potential energy as a function of Monte Carlo steps in the (a) early, (b) intermediate, and (c) final converged stages of a bicrystal sample containing two Σ5 (013) grain boundaries. Different limits of the Y-axes exist in each frame.

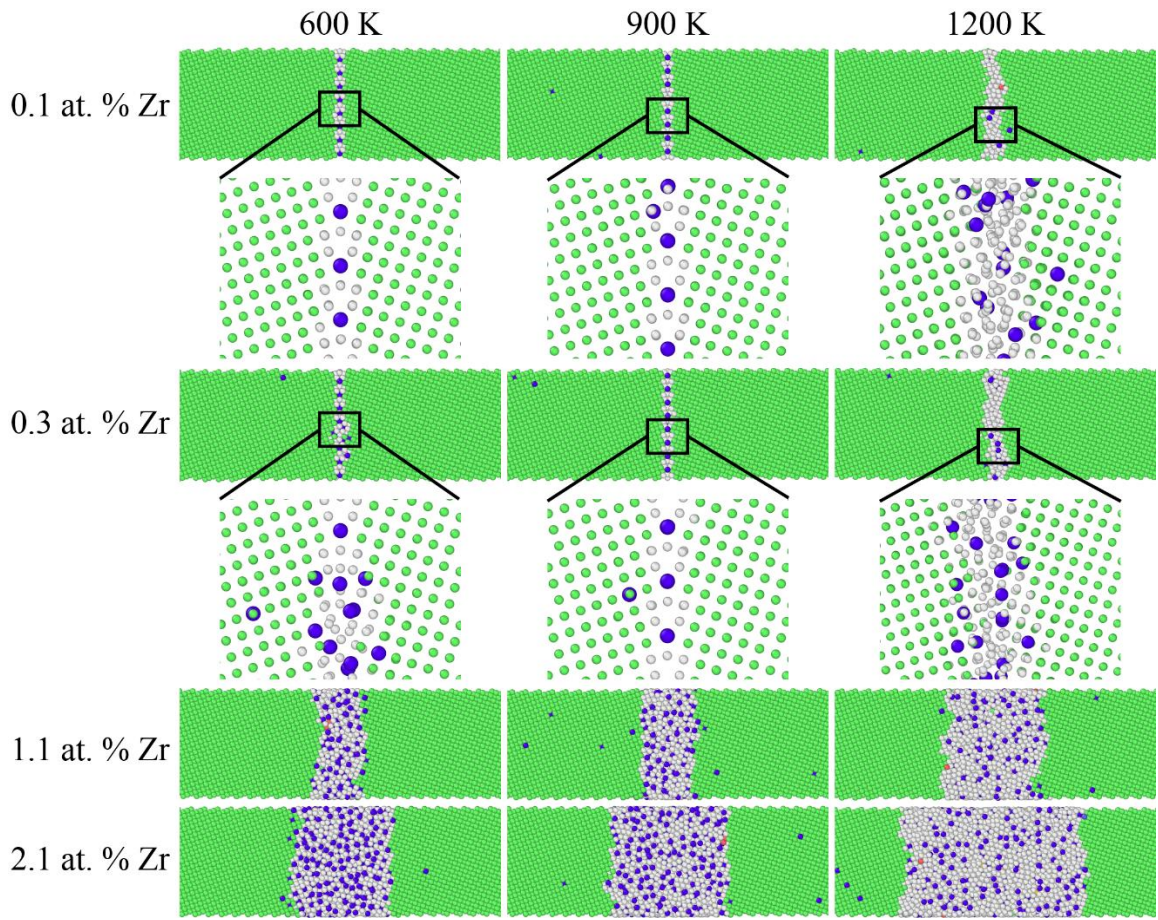

FIG. 3. Equilibrium structures of the Σ5 (013) grain boundary during the doping process at three temperatures. With increasing global composition, a transition from ordered grain boundaries with a single layer doping pattern to disordered intergranular films is observed.

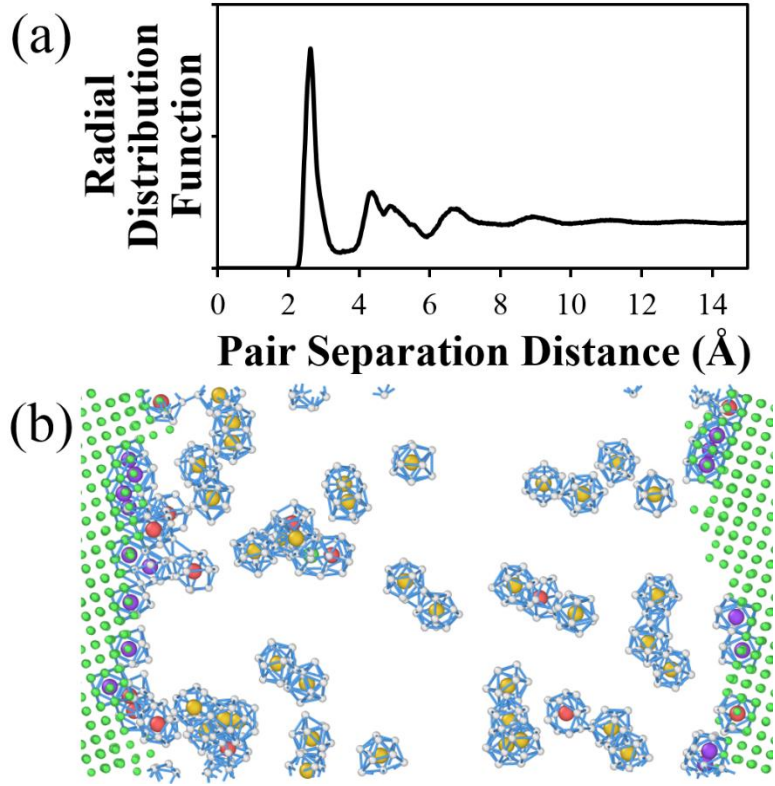

FIG. 4. (a) Radial distribution function of a disordered grain boundary region in a Σ5 (013) boundary at 1200 K and a global composition of 2.1 at. % Zr. No long-range, crystalline order is found. (b) The distribution and packing of local polyhedral clusters that contribute to limited short to medium-range order inside the disordered grain boundary region. Face centered cubic atoms are colored green, hexagonal close packed atoms are colored red, body centered cubic atoms are colored purple, and icosahedral atoms appear as yellow. Nearest neighbor atoms appear as vertices of the polyhedra.

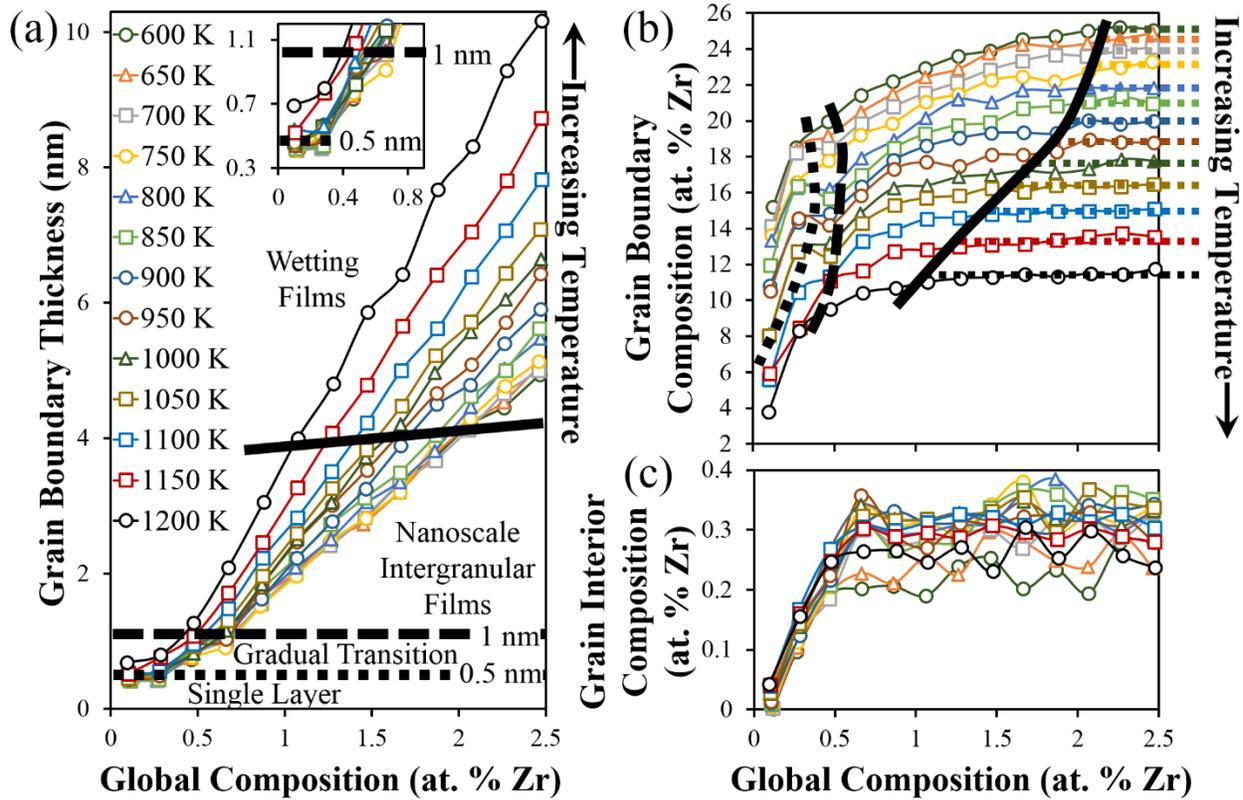

FIG. 5. (a) Grain boundary thickness, (b) grain boundary composition, and (c) grain interior composition for the sample containing Σ5 (013) grain boundaries as a function of global composition for different doping temperatures, indicating a gradual premelting transition. Inset to (a) shows the grain boundary thickness at the beginning of the doping process. The thick dotted, dashed, and solid lines (b) mark the boundaries between ordered grain boundary, gradual transition region, nanoscale intergranular film, and wetting film. The corresponding lines in (a) mark the approximate grain boundary thickness at the boundaries. Horizontal dotted lines in (b) mark the equilibrium grain boundary composition of the wetting film formed at temperatures denoted by the line color. The data at global compositions between 2.5 and 4.0 at. % Zr follows the same plateau in (b) and (c), and is therefore not shown here.

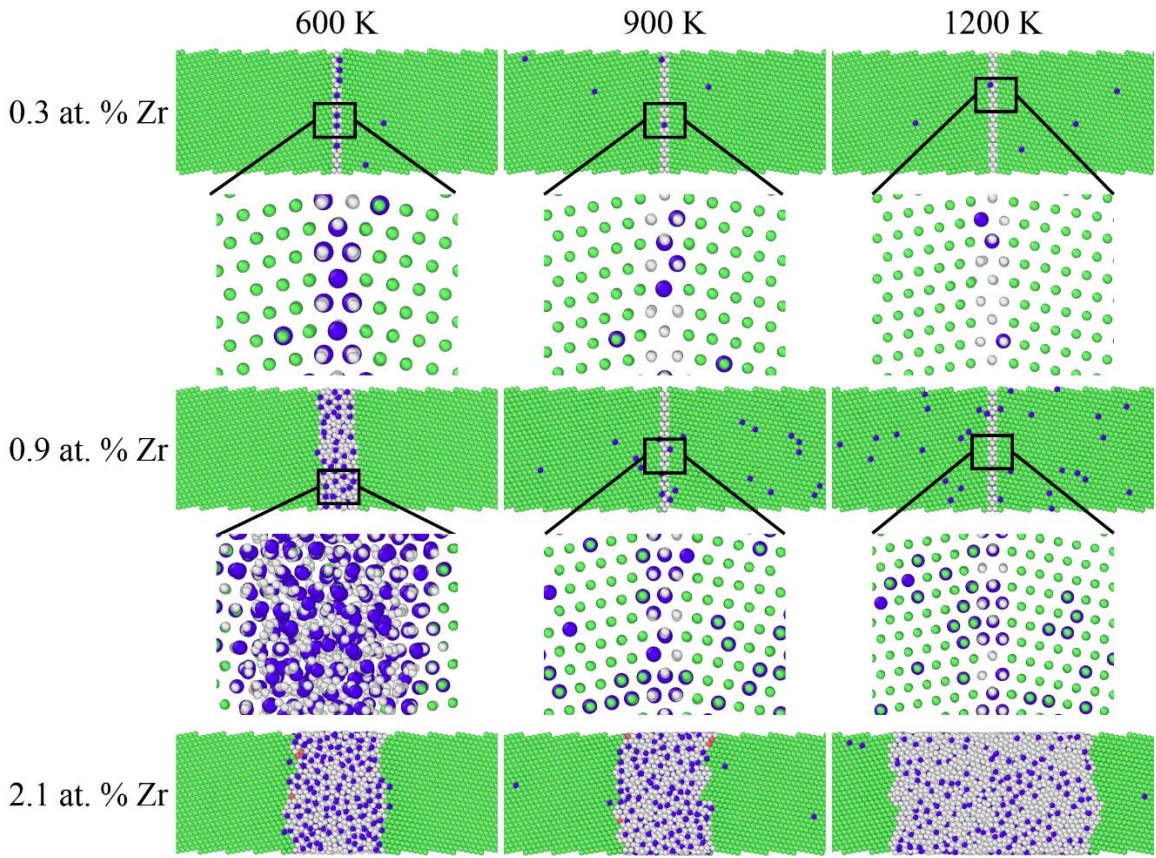

FIG. 6. Equilibrium structures of the Σ11 (113) grain boundary during the doping process at three temperatures. With increasing global composition, the transition from an ordered grain boundary to a disordered film first occurs at 600 K, then at 1200 K and 900 K.

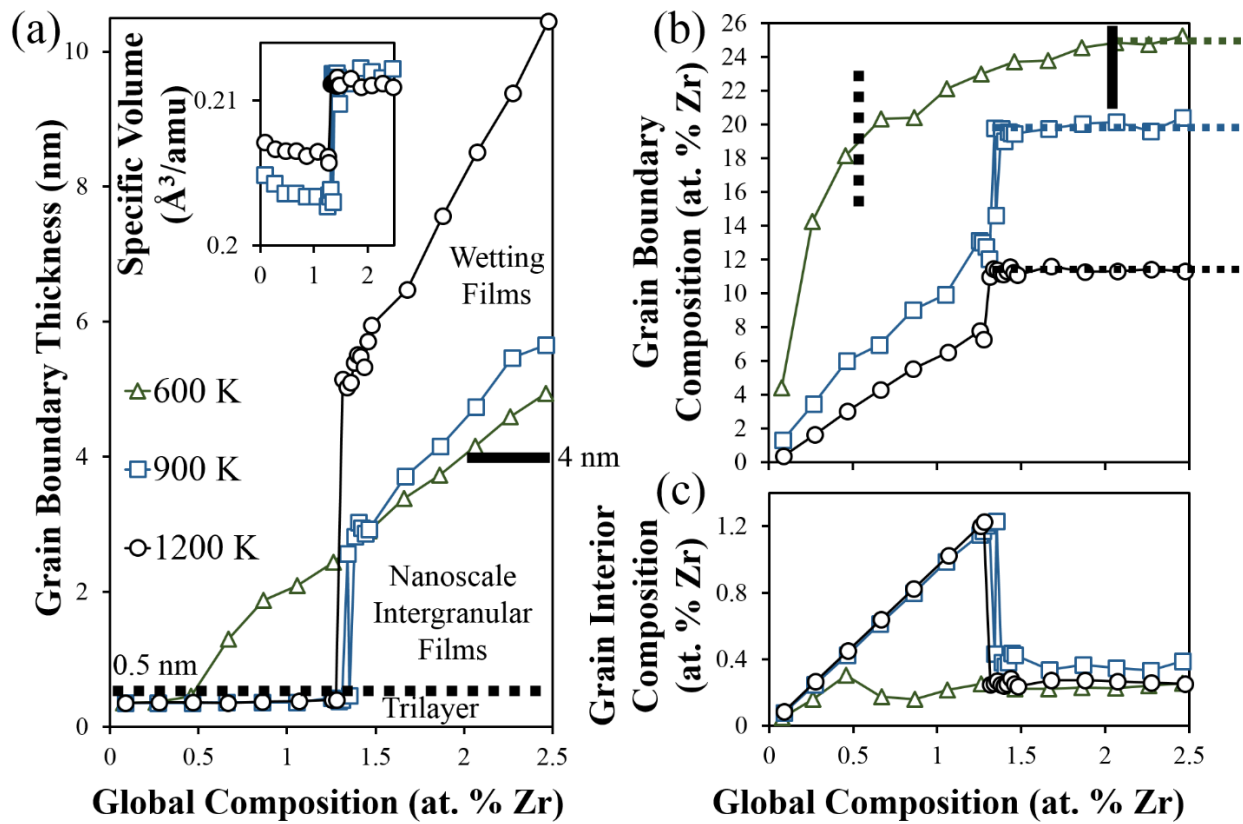

FIG. 7. (a) Grain boundary thickness, (b) grain boundary composition, and (c) grain interior composition for the sample containing Σ11 (113) grain boundaries as a function of global composition at three doping temperatures. The specific volume (volume per atomic mass unit) of the grain boundary at 900 K and 1200 K is also shown as an inset to (a). At 600 K, the boundaries between four doping stages are marked with thick lines. The data at global compositions between 2.5 and 4.0 at. % Zr follows the same plateau in (b) and (c), and is therefore not shown here.

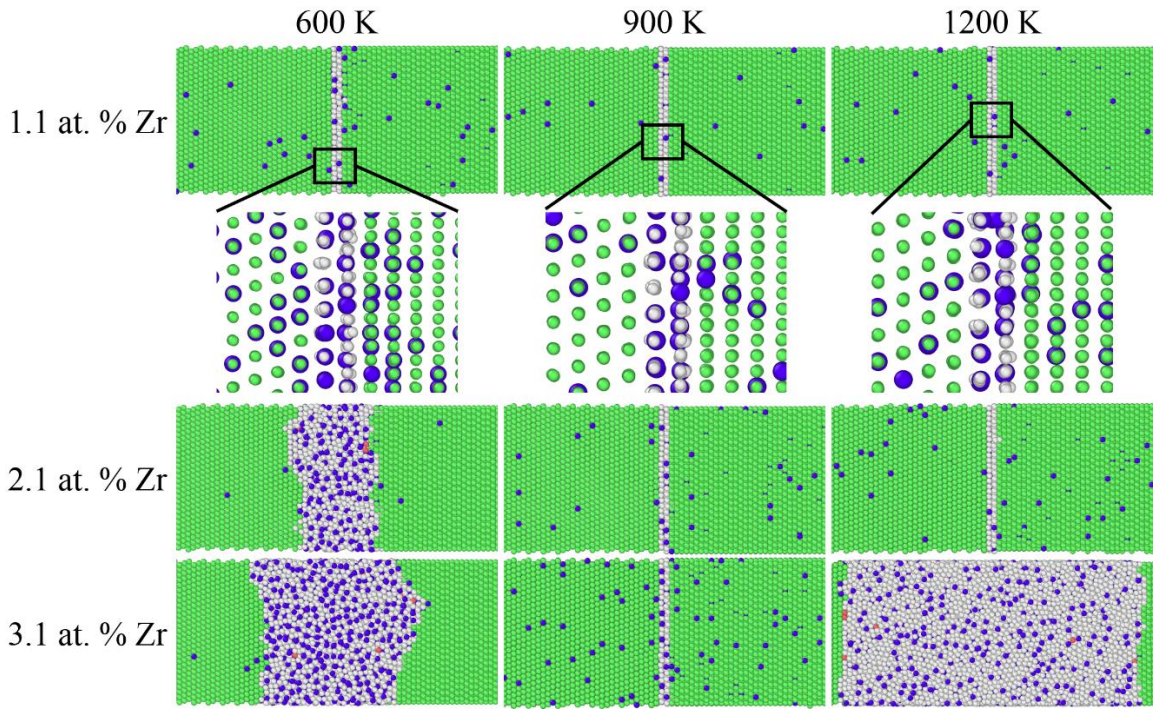

FIG. 8. Equilibrium structures of the twist (111) 30º grain boundary during the doping process at three temperatures. With increasing global composition, the transition from an ordered grain boundary to a disordered film first occurs at 600 K, then at 1200 K. No transition occurs at 900 K for the entire range of global composition studied (up to 4 at. % Zr).

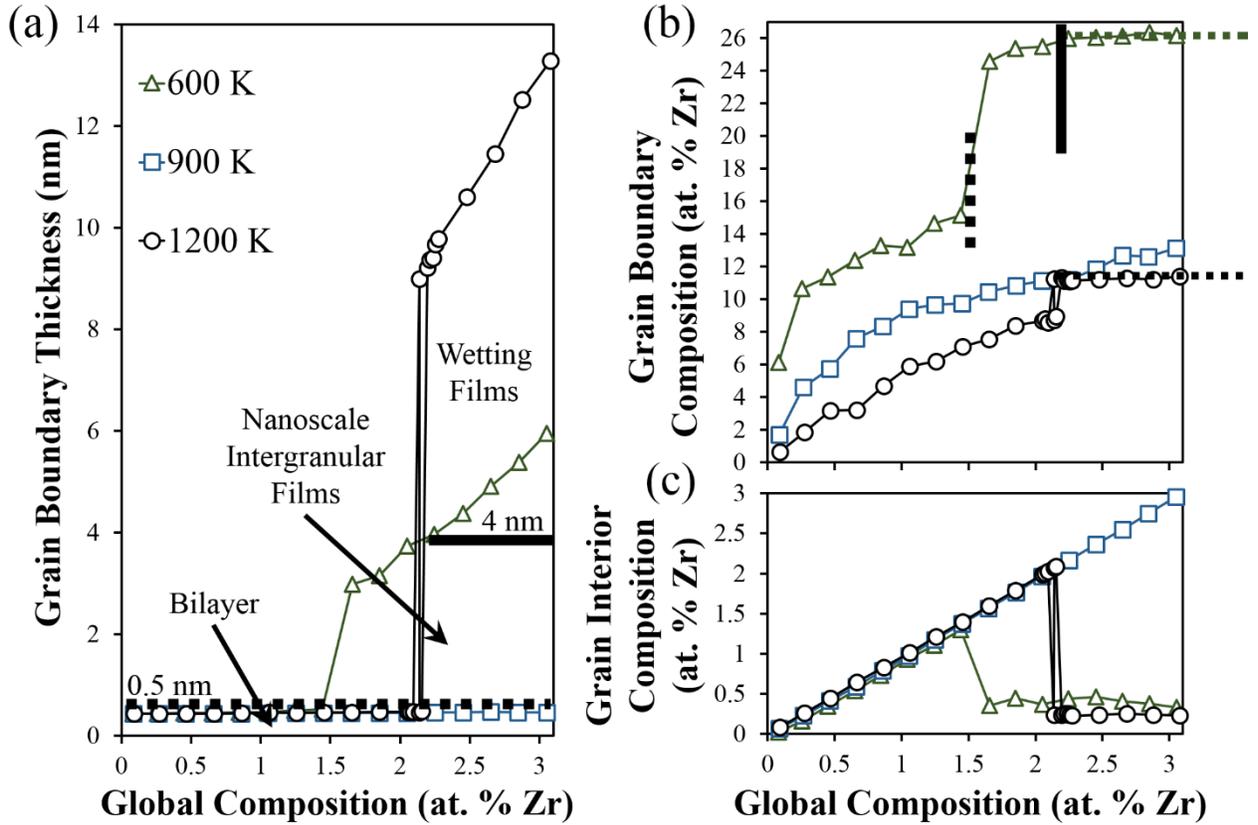

FIG. 9. (a) Grain boundary thickness, (b) grain boundary composition, and (c) grain interior composition for the sample containing twist (111) 30º grain boundaries as a function of global composition at three doping temperatures, indicating an abrupt transition at 600 K and 1200 K, but no transition at 900 K. At 600 K, the thick lines in (b) mark the boundaries between different doping stages and thick lines in (a) mark the corresponding approximate grain boundary thickness. The data at global compositions between 3.0 and 4.0 at. % Zr is not shown here, but follows the same trends.

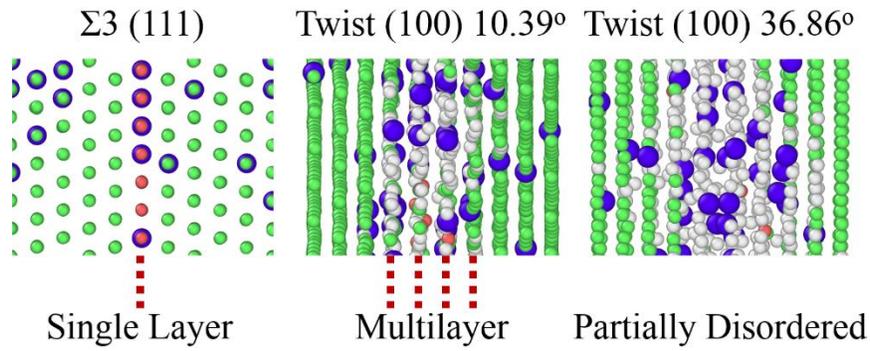

FIG. 10. Single layer, multilayer, and partially disordered equilibrium structures observed in the Σ3 (111) twin boundary, twist (100) 10.39° and twist (100) 36.86° grain boundaries at 1200 K at global compositions of 1.46, 0.86, and 0.47 at. % Zr, respectively.

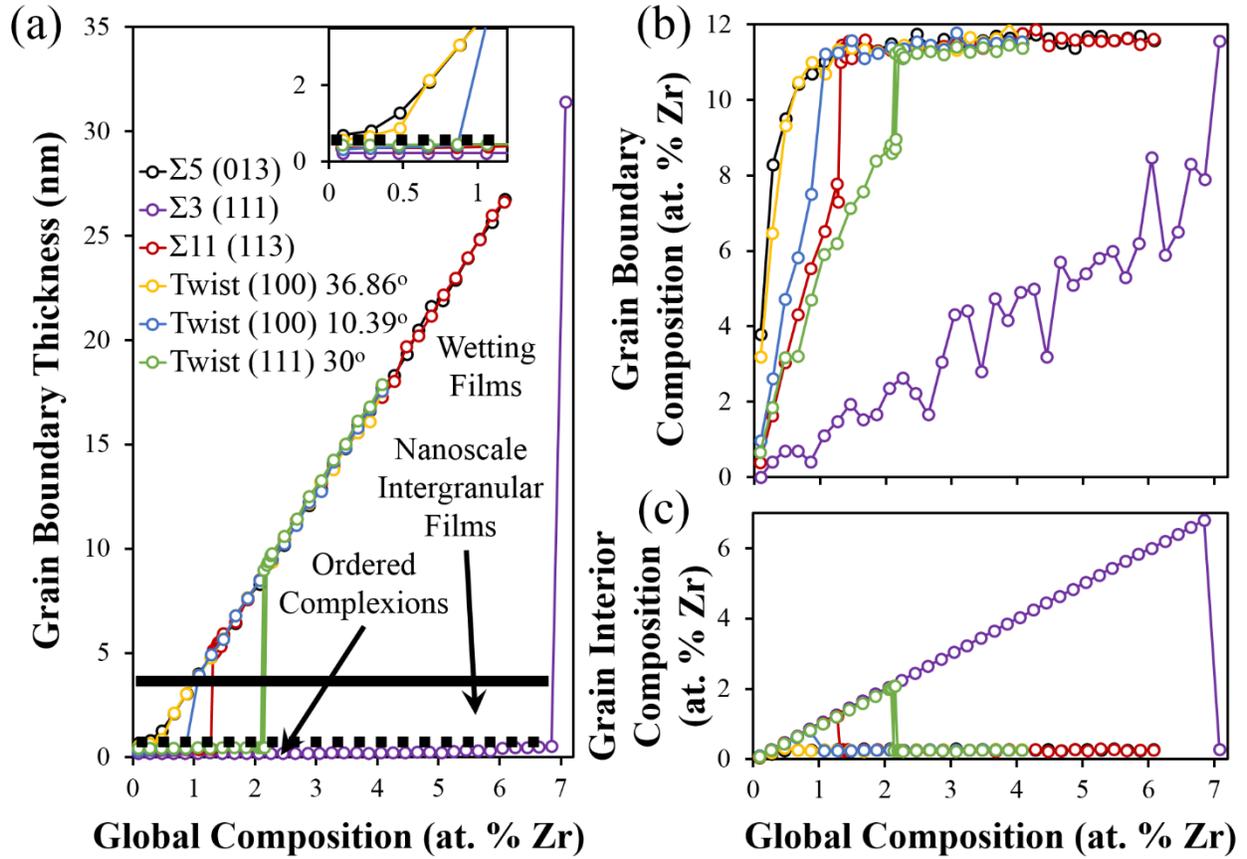

FIG. 11. (a) Grain boundary thickness, (b) grain boundary composition, and (c) grain interior composition for all the samples as a function of global composition at temperature of 1200 K, indicating gradual disordering transitions for Σ5 (013) and twist (100) 36.86º grain boundaries but abrupt transitions at the rest of grain boundaries. Inset to (a) shows the grain boundary thickness at the beginning of the doping process.

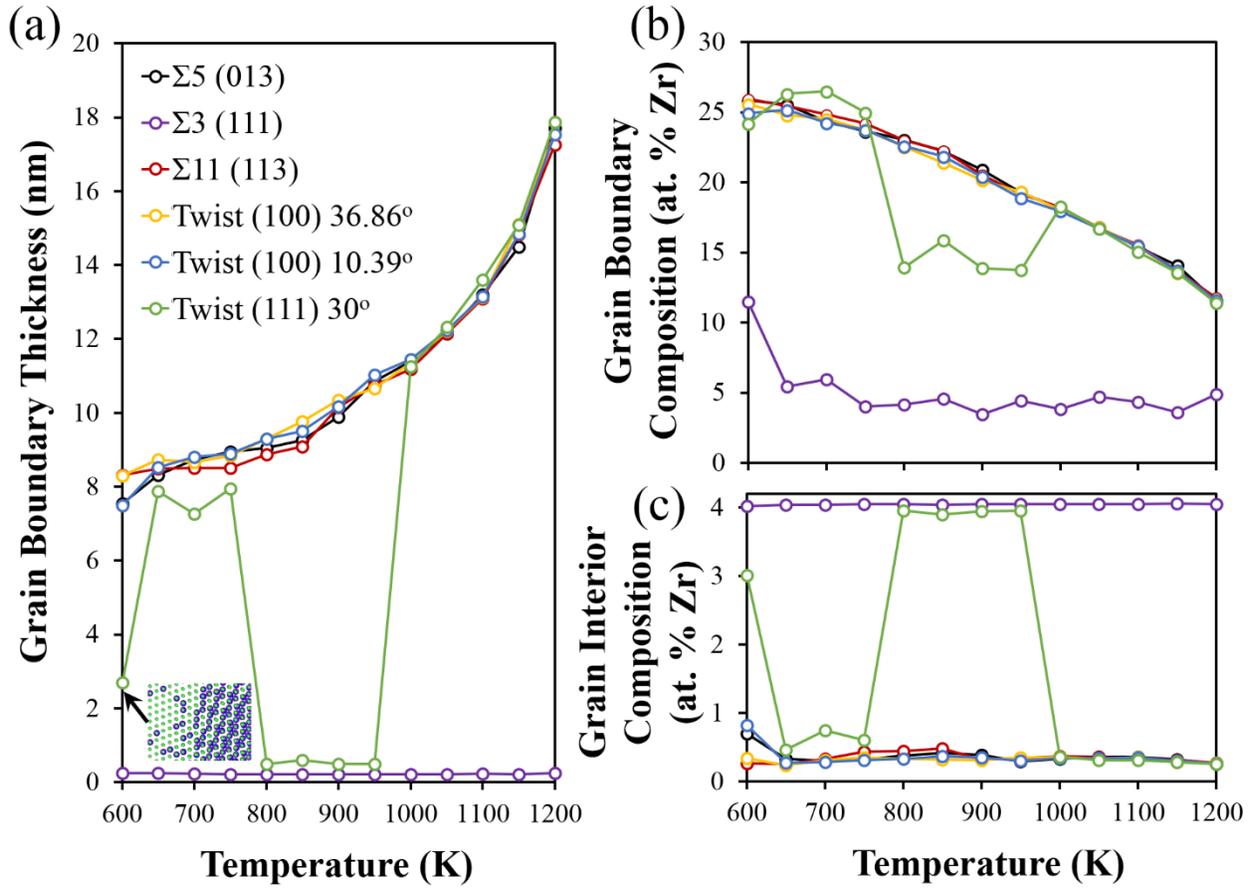

FIG. 12. (a) Grain boundary thickness, (b) grain boundary composition, and (c) grain interior composition for all the samples as a function of temperature while fixing the global composition to 4.0 at. % Zr. The disordering transition at twist (111) 30° grain boundary only occurs at low and high temperatures, but not at intermediate temperatures. The inset to (a) shows the second crystalline phase formed in the twist (111) 30° boundary at 600 K.

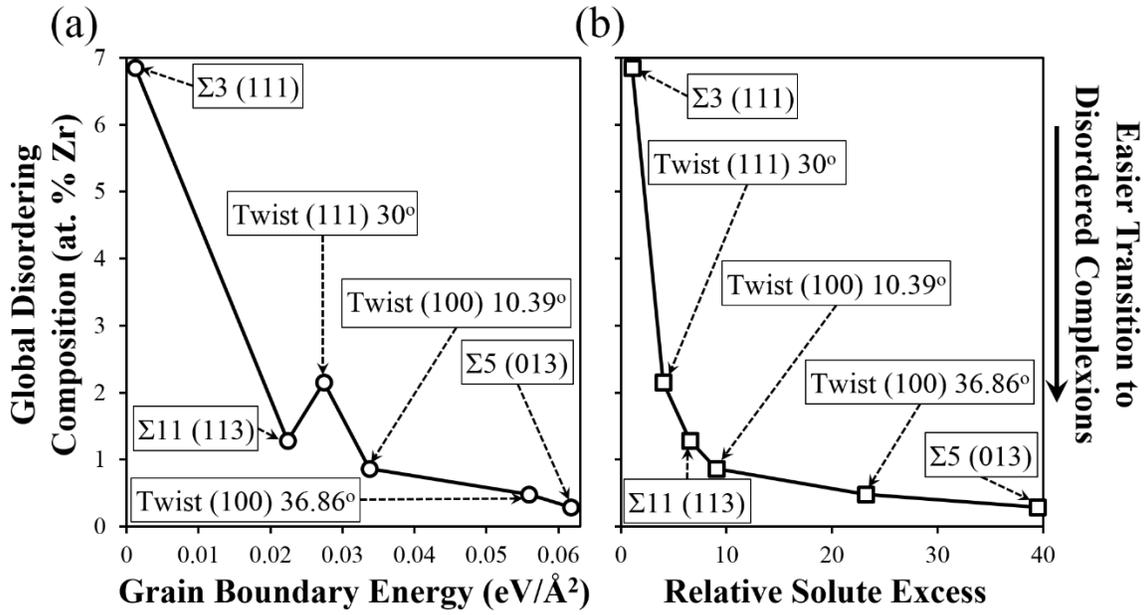

FIG. 13. Global disordering composition at transition at 1200 K as a function of (a) grain boundary energy and (b) relative solute excess (the ratio of grain boundary composition divided by grain interior composition) which gives a more consistent indicator for the tendency for segregation-induced disordering transitions.

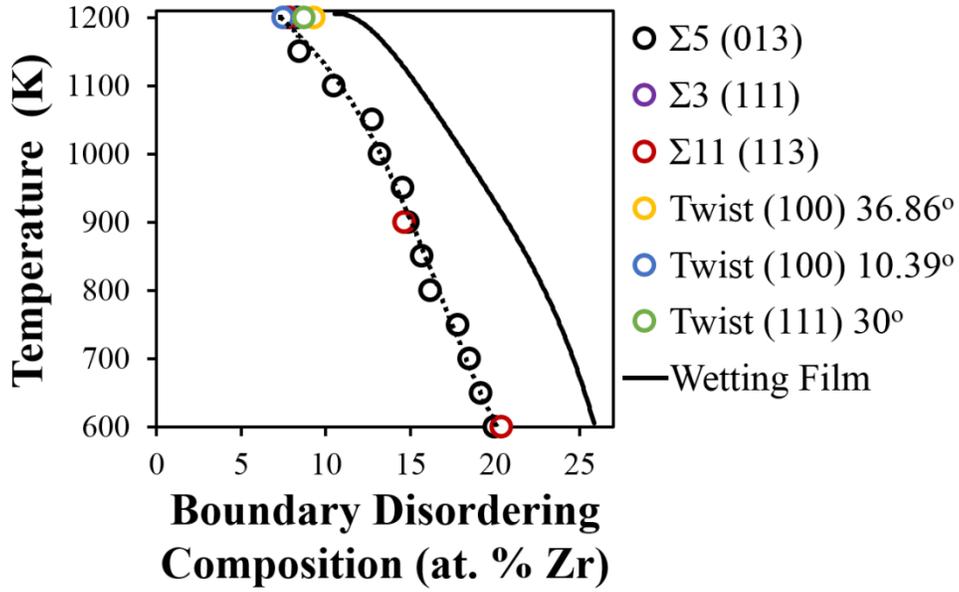

FIG. 14. Boundary disordering composition, or the critical grain boundary composition above which disordered intergranular films form, at different temperatures. The equilibrium concentration of disordered wetting films at different temperatures are also plotted, with data retrieved from Fig. 12(b) for the Σ5 (013) grain boundary.

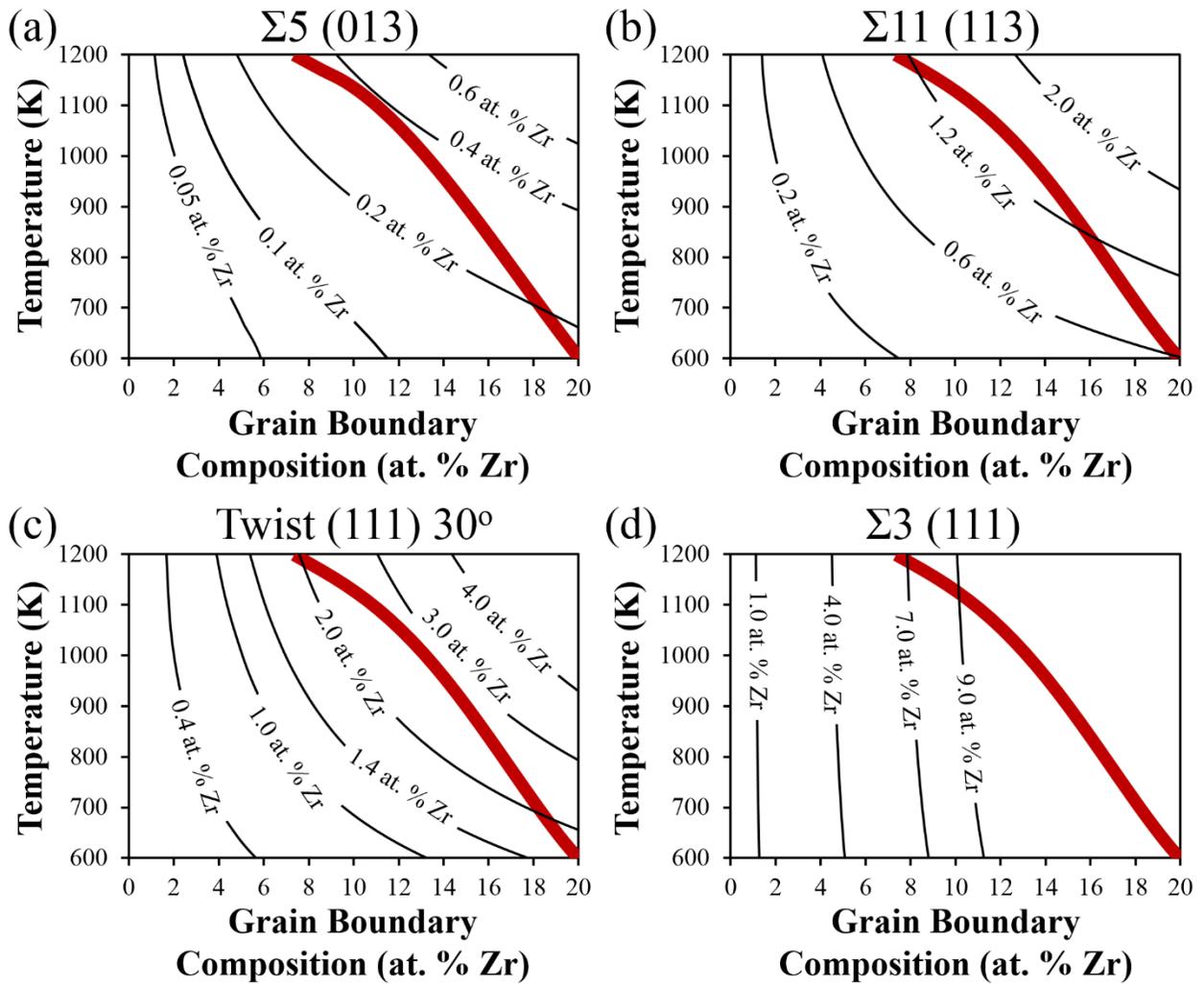

FIG. 15. Disordering transition diagrams based on Langmuir-McLean model where the segregation energy is adjusted so that the resulted relative solute excess is close to that of (a) Σ5 (013), (b) Σ11 (113), (c) Twist (111) 30º, and (d) Σ3 (111) grain boundaries. The black solid curves were plotted according to the Langmuir-McLean model and the red thick line is the fitted critical value in Fig. 14, above which disordering transitions occur.